\documentclass[runningheads]{llncs}
\usepackage[T1]{fontenc}
\usepackage{graphicx,wrapfig}
\usepackage[linesnumbered,ruled]{algorithm2e}
\usepackage{algorithmic}

\usepackage{graphicx}
\usepackage[colorlinks=true, allcolors=blue]{hyperref}
\usepackage{tikz-cd}
\usetikzlibrary{patterns}
\usepackage{subfigure}
\usepackage{latexsym}
\usepackage[flushleft]{threeparttable}
\usepackage{adjustbox}
\usepackage{epstopdf}
\usepackage{gensymb,xspace}
\usepackage{multicol,multirow,makecell}
\usepackage{stmaryrd}

\usepackage{booktabs} 

\newcommand{\R}{\ensuremath{\mathbb{R}}\xspace}

\makeatletter
\newcommand{\algorithmfootnote}[2][\footnotesize]{%
  \let\old@algocf@finish\@algocf@finish
  \def\@algocf@finish{\old@algocf@finish
    \leavevmode\rlap{\begin{minipage}{\linewidth}
    #1#2
    \end{minipage}}%
  }%
}

\usepackage{etoolbox}

\makeatletter
\g@addto@macro\@algocf@start{\vspace{0pt}}
\g@addto@macro\@algocf@finish{\vspace{-2pt}}
\makeatother

\usepackage{amsmath}
\usepackage{amssymb}

\DeclareMathOperator*{\argmin}{arg\,min}

\begin{document}
\title{Identification of Cellular Automata on Spaces of Bernoulli Probability Measures}
\titlerunning{Identification of Cellular Automata on Bernoulli Probability Measures}

\author{Faizal Hafiz\inst{1} \and
Amelia Kunze\inst{2} \and
Enrico Formenti \inst{2} \and
Davide La Torre\inst{1}}
\authorrunning{Hafiz et al.}

\institute{SKEMA Business School, Sophia Antipolis, France\\
\email{\{faizal.hafiz, davide.latorre\}@skema.edu}\\
 \and Universit\'{e} C\^{o}te d'Azur, CNRS, i3S, Nice, France\\
\email{\{amelia.kunze, enrico.formenti\}@univ-cotedazur.fr}
}
\maketitle              
\begin{abstract}
Classical Cellular Automata (CCAs) are a powerful computational framework for modeling global spatio-temporal dynamics with local interactions. While CCAs have been applied across numerous scientific fields, identifying the local rule that governs observed dynamics remains a challenging task. Moreover, the underlying assumption of deterministic cell states often limits the applicability of CCAs to systems characterized by inherent uncertainty. This study, therefore, focuses on the identification of Cellular Automata on spaces of probability measures (CAMs), where cell states are represented by probability distributions. This framework enables the modeling of systems with probabilistic uncertainty and spatially varying dynamics. Moreover, we formulate the local rule identification problem as a parameter estimation problem and propose a meta-heuristic search based on Self-adaptive Differential Evolution (SaDE) to estimate local rule parameters accurately from the observed data. The efficacy of the proposed approach is demonstrated through local rule identification in two-dimensional CAMs with varying neighborhood types and radii. 


\end{abstract}%

\section{Introduction}

Classical Cellular Automata (CCAs) have long been a cornerstone computational framework for modeling and understanding complex systems driven by local interactions. Their simplicity, rooted in a finite set of states and uniform local rules, belies their ability to produce rich and diverse dynamical behaviors. However, a fundamental limitation of CCAs lies in their assumption of complete certainty in the state of all cells. In many real-world systems, uncertainty is inherent in the dynamics, and this deterministic assumption restricts the ability of CCAs to model such dynamics effectively.

To address this limitation, we propose Cellular Automata on spaces of Ber\-noul\-li probability Measures (CAMs), a novel generalization of CCAs that incorporates probabilistic uncertainty into the framework. The proposed framework is essentially a special case of the recently proposed Cellular Automata on Measures by the authors in~\cite{Lillepreprint}. In our setting, the state of each cell is represented by a Bernoulli probability measure, $\mu(x)\in[0,1]$, rather than a discrete value, $x\in\{0,1\}$, and local rules operate on configurations of  Bernoulli probability measures. This extension allows for the modeling of systems with spatially varying probabilities and inherent randomness, significantly broadening the scope of cellular automata to mimic complex and realistic phenomena.\vspace{-.05cm}

Cellular automata were first proposed as an idealized biological system with the aim of modeling biological self-reproduction \cite{Delorme1999}. They have since been employed as tools in numerous application areas including neural networks, cryptography, and quantum computing (e.g., \cite{FarinaD08,Gilpin2019,Formenti2014,Farrelly2020}).
Cellular automata have a particularly rich history as discrete models of molecular dynamics 
for reaction-diffusion equations and fluid dynamics \cite{Wolfram1986}.
These types of applications yield cellular automata which can model the spread of the pollution or the spread of disease in a population. Given this, our work presented below offers a means to successfully construct a cellular automaton which can address meaningful real-world problems. 

This study, in particular, focuses on the \emph{inverse} problem of determining the local rule of CAMs, which governs the observed spatio-temporal patterns. This inference process is often referred to as the \emph{identification} of local rules~\cite{Adamatzky:2018,zhao:Billings:2007}, and it represents a foundational step in modeling real-world systems~\cite{Hafiz:Swain:Flaoting:2019,Hafiz:Swain:2019}.
The existing investigations often approach the CCA identification as a two-step process, which involves determining the approximate \emph{neighborhood size} followed by further refinements. For instance, Yang and Billings~\cite{Yang:Billings:2000} proposed neighborhood detection by selecting cells based on their \emph{contribution} to the updated cell. Often, combinations of candidate neighborhood cells and the updated cell are treated as \emph{patterns}~\cite{Mei:Billings:2005,Zhao:Billings:2006} or \emph{evidences}~\cite{Maeda:Sakama:2007}, which are subsequently used to provide statistics or heuristics to select the appropriate neighborhood. The next identification step involves further refinements, such as removing \emph{redundant} cells~\cite{Maeda:Sakama:2007} and rule refinement using Genetic Algorithms~\cite{Yang:Billings:2000}. We refer to~\cite{Adamatzky:2018,zhao:Billings:2007} and the references therein for a detailed treatment of this topic.

It is worth emphasizing that extending the existing CCA-focused identification approaches to CAMs, where a Bernoulli probability measure represents each cell state, may not be trivial. To this end, we demonstrate that the local rule of CAMs can be formulated as a convex combination of neighborhood cell states, which enables us to cast the identification task as an estimation of local rule parameters. Next, we propose a meta-heuristic search approach based on Self-adaptive Differential Evolution (SaDE)~\cite{Qin:Huang:2009}. This approach models the local rule parameter estimation as a constrained optimization problem to minimize the \emph{difference} between the observed and simulated states. We show that the distance may be easily computed for Bernoulli probability measures using the Monge-Kantorovich metric. This finding is leveraged to quantify the \emph{difference} between the observed and simulated CAM states, thus driving the optimization process. 

The rest of this article is organized as follows: we begin with a discussion of key properties of CCA in Section~\ref{sec:CCAproperties}, followed by the space of probability measures in Section~\ref{sec:probmeas} and the framework of Cellular Automata on Bernoulli Probability Measures in Section~\ref{sec:CAM}. The proposed formulation of CAM identification and the meta-heuristic search approach are provided in Section~\ref{sec:IP}. Next, different identification test scenarios and the corresponding results are discussed in Section~\ref{sec:Res}. This is followed by the conclusions in Section~\ref{sec:conclusions}.

\section{Classical Cellular Automata: Definitions and Main Properties}
\label{sec:CCAproperties}

A cellular automaton consists of a uniform grid of cells, usually infinite in extent. Each cell in the grid takes a value from a discrete set of possible \textit{states}. A particularly simple and classical case is that of binary states, in which every cell in the automaton is either ``alive'' (in which the state takes value $1$) or ``dead'' (in which the state takes value $0$). The automaton evolves at discrete time steps through synchronous updates to the state of all cells according to a local, space-invariant \textit{update rule}. 

\begin{wrapfigure}[20]{r}{0.33\textwidth}
\vspace{-\baselineskip} 
\centering
\begin{tikzpicture}[scale=0.45]
\tikzset{active/.style={pattern=north east lines, pattern color=black!90}}
\tikzset{center/.style={pattern=dots, pattern color=black}}

\foreach \x in {0,1,2,3,4} {
    \foreach \y in {6,7,8,9,10} {
        \fill[white] (\x, \y) rectangle ++(1,1);
        \draw (\x, \y) rectangle ++(1,1);
    }
}
\fill[active] (2,6) rectangle ++(1,1);
\fill[active] (1,7) rectangle ++(1,1);
\fill[active] (2,7) rectangle ++(1,1);
\fill[active] (3,7) rectangle ++(1,1);
\fill[active] (0,8) rectangle ++(1,1);
\fill[active] (1,8) rectangle ++(1,1);
\fill[center] (2,8) rectangle ++(1,1);
\fill[active] (3,8) rectangle ++(1,1);
\fill[active] (4,8) rectangle ++(1,1);
\fill[active] (1,9) rectangle ++(1,1);
\fill[active] (2,9) rectangle ++(1,1);
\fill[active] (3,9) rectangle ++(1,1);
\fill[active] (2,10) rectangle ++(1,1);
\node at (2.5,5.4) {\tiny Manhattan};

\foreach \x in {0,1,2,3,4} {
    \foreach \y in {0,1,2,3,4} {
        \fill[white] (\x, \y) rectangle ++(1,1);
        \draw (\x, \y) rectangle ++(1,1);
    }
}
\foreach \x in {0,1,2,3,4} {
    \foreach \y in {0,1,2,3,4} {
        \ifnum\x=2\relax\ifnum\y=2\relax\else\fill[active] (\x,\y) rectangle ++(1,1);\fi\else\fill[active] (\x,\y) rectangle ++(1,1);\fi
    }
}
\fill[white] (2,2) rectangle ++(1,1);
\fill[center] (2,2) rectangle ++(1,1);
\node at (2.5,-0.6) {\tiny Moore};
\end{tikzpicture}
\caption{Illustration of different neighborhoods in two-dimensional cellular automaton for the neighborhood radius, $r=2$.}
\label{fig:neighborhoods}
\vspace{-\baselineskip} 
\end{wrapfigure}
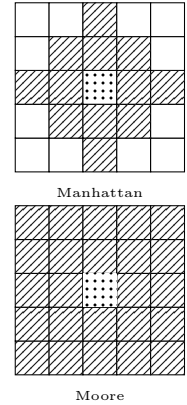

\subsection{Formal Definitions}

The cellular grid can be more formally described as a $d$-dimensional, $d \in \mathbb{N}$, regular lattice $\Lambda$. The finite set of states, which indicates all possible values that each cell in the lattice can take, is denoted by $\mathcal{S}$. A \textit{configuration} $x \in \mathcal{S}^\Lambda$ is a mapping that assigns a state $x_i \in \mathcal{S}$ to each cell $i \in \Lambda$.  Cells update their state by gathering information about the states of their neighbors. The \textit{neighborhood} $N_r(i)$ of a cell $i \in \Lambda$ is the closed ball of radius $r$ centered at $i$. The \textit{local rule} is a function $f: \mathcal{S}^{\lvert N_r(i) \rvert} \to \mathcal{S}$ that combines the neighboring states to generate local updates. A \textit{classical cellular automaton} (CCA) is then formally described by the structure $\langle d, \mathcal{S}, r, f \rangle$.

In this work, we focus on two-dimensional cellular automata with lattice $\Lambda = \mathbb{Z}^2$. Different neighborhoods can be obtained by changing how one measures distance from the central cell. We use two different neighborhoods which correspond to using the classical Manhattan and Moore distances:
\begin{small}    
\begin{align}
    \label{eq:top}
        N_r(i) & = \big\{ j \in \Lambda \ | \ d_1(i,j) \leq r \big\} \quad \texttt{Manhattan} \nonumber\\
        N_r(i) & = \big\{ j \in \Lambda \ | \ d_\infty(i,j) \leq r \big\} \quad \texttt{Moore}
\end{align}
\end{small}
where, $d_1$ and $d_\infty$ respectively denote Manhattan and Chebyshev distance. Note that the Manhattan neighborhood is quite often referred to as the Von Neumann neighborhood in the CCA literature. The size of neighborhood, $K$, is controlled by the radius $r$ and it is given by: $K=(2r^2+2r+1)$ (Manhattan) and $K = (2r+1)^2$ (Moore). The geometry of both neighborhood types is illustrated for $r=2$ in Fig.~\ref{fig:neighborhoods}.

In order to study CCA as dynamical systems, the set of states $\mathcal{S}$ is equipped with the topology induced by the discrete metric $d_D$ defined as follows
\begin{small}    
\begin{equation}
d_D(s, t) =
\begin{cases}
1 & \text{if } s \neq t, \\
0 & \text{if } s = t,
\end{cases}
\end{equation}
\end{small}
for all $s,t\in \mathcal{S}$. Then, 
$\mathcal{S}^\Lambda$ is endowed with \emph{Cantor topology}, \textit{i.e.}, 
the standard product
topology induced by the discrete topology on $\mathcal{S}$. Consider the \emph{Cantor distance} defined as
\begin{small}    
\begin{equation}
d_C(x, y) = \sum_{i \in \Lambda} \frac{d_D(x_i, y_i)}{s^{|i|}},
\end{equation}
\end{small}
where $s$ is the size of $\mathcal{S}$
(recall that $\mathcal{S}$ is a finite set here). It is well-known that the Cantor distance induces the Cantor topology on $\mathcal{S}^\Lambda$.

For a fixed $j\in \mathcal{S}$, the \emph{shift map} $\sigma_j$ is a very well-known and widely studied CA defined as follows 
\begin{equation}
\sigma_j(x)_i = x_{i+j}, \quad \forall x \in \mathcal{S}^\Lambda, \ \forall i \in \Lambda.
\end{equation}

The famous Curtis-Hedlund-Lyndon theorem established that CCAs are exactly those continuous maps commuting with the shift \cite{hedlund69}.
For an up-to-date bibliography on CAs and their variants, see, for instance~\cite{CattaneoDM02,CattaneoDM04,Kari05,BhattacharjeeNR20}.

In the next section, we introduce a metric for probability measures to be used in our CAM setting and provide a useful result. 

\section{The Space of Probability Measures}
\label{sec:probmeas}

In the following let us suppose that the state space $S = \{0,1\}$ is endowed with a metric $d$ and that $(S,d)$ is a compact metric space. Let $\mathcal{B}(S)$ be the Borel sigma-algebra defined on $S$. Let us denote by $\mathcal{M}(S)$ the set of all probability measures defined on $S$. $\mathcal{M}(S)$ can be endowed with the Monge-Kantorovich metric defined as
\begin{small}    
\begin{equation}
d_{MK}(\mu,\nu) = \sup_{g\in Lip_1(S)} \int_S g d\mu - \int_S g d\nu     
\end{equation}
\end{small}
where $Lip_1(S)$ is the set of all 1-Lipschitz functions on $S$, that is 
\begin{small}
\begin{equation}
Lip_1(S) = \{g:S\to\R\text{ s.\@\,t.\@\;} |g(z)-g(y)|\le d(x,y)\}.   
\end{equation}
\end{small}

The Monge-Kantorovic metric arose from a classical problem in transportation of mass 
\cite{Formentipreprint}. Although it is often used to measure the distance between probability measures, its computation may not appear intuitive. The following result shows that the Monge-Kantorovich metric simplifies in the case of Bernoulli probability measures. 

Let $\mu$ be a Bernoulli probability measure with parameter $p$, i.e., $\mu(z)=p^z(1-p)^{(1-z)}$ for $z \in \{0,1\}$. Let $\nu$ be a Bernoulli probability measure with parameter $q$, i.e., $\nu(z)=q^z(1-q)^{(1-z)}$ for $z \in \{0,1\}$.

For the Bernoulli probability measure $\mu$ and an arbitrary function $f \in Lip_1(S)$ we have
\begin{small}    
\begin{equation}
    \int_S gd\mu = E[g(x)] = g(1)\mu(1) + g(0)\mu(0)
\end{equation}
\end{small}
with the analogous equation holding for $\nu$. Their difference is 
\begin{small}\begin{align}
    \int_S gd\mu - \int_S gd\nu &= g(1)\mu(1) + g(0)\mu(0) - (g(1)\nu(1) + g(0)\nu(0))\nonumber\\
    &= g(1)p + g(0)(1-p) - g(1)q - g(0)(1-q)\nonumber\\
    &= (g(1) - g(0))(p-q).
    \label{eqn:computedmk}
\end{align}
\end{small}
For any $g \in Lip_1(S)$ we have $\lvert g(1)-g(0) \rvert \leq d(1,0)=1$ and therefore $\sup_{g\in Lip_1(S)} \lvert g(1) - g(0)\rvert = 1$. 
In addition, if $g \in Lip_1(S)$ then $-g \in Lip_1(S)$ giving
\begin{small}    
\begin{align}
d_{MK}(\mu, \nu) &= \sup_{g \in Lip_1(S)} \left\lvert 
  \int_S gd\mu  - \int_S gd\nu \right\rvert = \sup_{g\in Lip_1(S)} \lvert g(1)-g(0) \rvert \lvert p-q \rvert \nonumber\\
 & = \lvert p-q \rvert.
    \label{eqn:dMKsimplify}
\end{align}
\end{small}

\section{Cellular Automata on spaces of Bernoulli Probability Measures}
\label{sec:CAM}

In the following, we extend the classical cellular automaton with state space $\mathcal{S}=\{0,1\}$. We assume that there is some inherent randomness associated with the realization of the state $x_i=1$ in a given cell $i$. The probability of finding a state $x_j \in S$ in cell $i$ is thus described by a Bernoulli random process,
\begin{equation}
 \mu_i(x_j) = p^{x_j}(1-p)^{1-x_j} \qquad \text{ for } x_j \in S=\{0,1\}.
\end{equation}

In this framework, the state set $S$ is replaced with $\mathcal{M}(S)$, which denotes the set of Bernoulli probability measures defined on $S$. In each cell, the Bernoulli random process is described completely by the probability $p_i \in [0,1]$. The local rule thus operates on configurations of Bernoulli probability measures instead of operating directly on the set of states $S=\{0,1\}$. Then the dynamics generated by a spatially-varying Bernoulli distributions on the state space $S=\{0,1\}$ can be completely characterized by the evolution of the values $p_i \in [0,1]$ on the lattice $\Lambda$. In other words, our model is equivalent to a cellular automaton with continuous state space $S=[0,1]$. The local rule is then a function $f:[0,1]^{|\mathcal{N}|} \to [0,1]$ which updates the probability $p_{i}(t)$ to $p_{i}(t+1)$ based on the values of its neighbors $p_{j}(t) \in N_r(i)$.

We define a distance between two configurations $x,y \in \mathcal{M}(S)^\Lambda$ in the following way.
\begin{definition}
\label{def1}
For any pair of configurations \(x, y \in \mathcal{M}(S)^\Lambda\), the distance \(d_{\mathcal{M}}\) is defined as
\begin{equation}
\label{eq:dM}
d_{\mathcal{M}}(x, y)= 
\sum_{i \in \Lambda} \frac{d_{MK}(x_i, y_i)}{2^{|i|}}
= \sum_{i \in \Lambda} \frac{\lvert p_{x_i} - p_{y_i}\lvert} {2^{|i|}}.
\end{equation}
where $p_{x_i}$ and $p_{y_i}$ denote the Bernoulli parameters at the $i^\text{th}$ cells of the configurations $x$ and $y$, respectively, and \(|\cdot|\) denotes the norm of \(i\).
\end{definition}

In the above definition, we have used the fact that in the Bernoulli case, the Monge-Kantorovich metric simplifies to the $L^1$ distance between the parameters, as shown in Eq.~(\ref{eqn:dMKsimplify}). For Bernoulli probabilities, the metric $d_\mathcal{M}$ is a weighted distance which emphasizes agreement between the centrally-located cells in the two configurations.

Observe that the space of configurations $\mathcal{M}(S)^\Lambda$ is closed under convex combinations of its elements. For the proof, it is sufficient to show that the interval $[0,1]$ is a convex set. This fact motivates a particularly useful choice of local rule, denoted by $f$, which we first introduced and investigated in \cite{Lillepreprint}. In essence, the local map $f$ takes convex combinations of neighboring cells as explained below.

Let $\{\theta_k\}$ be a set of weights in $[0,1]$ such that $\sum_{k\in N_r(i)} \theta_k = 1$, for any cell $i \in \Lambda$ and the corresponding neighbors $k\in N_r(i)$, then the action of local rule $\delta_C$ on any $x\in \mathcal{M}(S)^\Lambda$ is given by,
\begin{small}    
\begin{align}
   \label{eqn:ccrule}
   f(x)_i & = \sum \limits_{k=1}^{K} \theta_k \mathcal{N}_i, \quad \text{where,} \quad \mathcal{N}_i = \left[x_{k} \mid k \in N_r(i)\right] 
\end{align}
\end{small}
$\mathcal{N}_i$ is the vector of neighborhood states of the $i^{th}-$cell; and $K$ gives the total number of neighbors. The vector $\mathcal{N}_i$ is formed by taking the element of the neighborhood in Marshall order. Note that the set of weights $\theta_k$ is only dependent on the position of the cell $k\in N_r(i)$ and is not dependent on cell $i$. 

It is worth highlighting that we introduced CAM as generic framework in~\cite{Lillepreprint}, where $\mathcal{M}(S)$ is not limited to Bernoulli measures and can represent any arbitrary spatially-varying probability measures. This framework includes the classical setting, if Dirac measures concentrated on specific states, $s \in S$, are employed. In this more general setting, the type of probability measure defined at each cell may also vary. In contrast, the present study deals only with the simple case of spatially-varying Bernoulli probability measures, which is enough for our interests relating to the identification problems presented below. We direct the reader to \cite{Lillepreprint} for a detailed treatment of CAMs.

\section{Inverse Problems for CAMs}
\label{sec:IP}

\subsection{Problem Formulation}
\label{subsec:IP}

Identification is the first step of the modeling process, typically involving model inference from observed system behavior. This is of particular interest in practical applications where the local rule governing the observed behavior is unknown. The identification of such an optimal rule is often non-trivial due to the associated exponential search region, and it is typically approached as an optimization problem~\cite{Adamatzky:2018,zhao:Billings:2007}. To understand this further, consider the proposed CAM framework, which can formally be described as the following structure: $\langle \Lambda, \mathcal{M}(S), N_r, \delta_C \rangle$. Here, $\Lambda$ denotes a $d-$dimensional cellular lattice, $\mathcal{M}(S)$ gives the Bernoulli probability measures defined on $S$, $N_r$ denotes the $r-$ radius neighborhood, and $\delta_C$ represents a local transition rule. At a particular time step $t$, each cell $i\in \Lambda$ evolves as the convex combination of its neighbors,
\begin{small}
\begin{align}
    \label{eqn:direct_problem}
    x_i(t+1) & = f(x(t))_i = \sum \limits_{k=1}^{K} \theta_k \mathcal{N}_i(t) \qquad \forall \ t=0, \ldots, T-1 
\end{align}
\end{small}
where, $\mathcal{N}_i(t)$ is the vector of neighborhood states of the $i^{th}$ cell at time $t$. Eq.~(\ref{eqn:direct_problem}) represents the \emph{direct problem}, which can be used to study emergent global behavior with a given local rule, \textit{i.e.}, when parameters $\{\theta_k\}$ are known.

In contrast, the identification (or inverse problem) aims to identify the local rule, $f(\cdot)$, from the observed spatio-temporal trajectories of a discrete dynamical system. This can be viewed as a two-step process: the first step involves the selection of an appropriate neighborhood type and radius, $N_r$, which is followed by the estimation of the corresponding parameters, $\{\theta_k\}$. To understand this further, consider the observed data $\mathcal{O}$ collected over $T$ time steps:
\begin{small}    
\begin{align}
    \label{eq:id_data}
    \mathcal{O} &= \bigcup_{t=0}^{T-1} \left\{\left(\mathcal{N}_i(t), x_i(t+1)\right) \mid i = 1,\ldots,|\Lambda|\right\}
\end{align}
\end{small}
As discussed earlier, the neighborhood type (\textit{i.e.}, Manhattan or Moore, see Fig.~\ref{fig:neighborhoods}) and radius $r$ determine the neighborhood size, and, thereby, the total number of parameters, $K$. This study assumes that such parameters are \emph{a priori} determined either through an information-theoretic (\textit{e.g.}, mutual information) or an empirical trial-and-error criterion~\cite{Adamatzky:2018,Maeda:Sakama:2007,Mei:Billings:2005,Yang:Billings:2000,Zhao:Billings:2006}. This \emph{a priori} step reduces the identification to a parameter estimation problem with the objective of minimizing the distance between the observed states ($x_i$) and the simulated states ($\hat{x}_i$) as follows:
\begin{small}    
\begin{align}
    \label{eq:IP}
    \Theta^\ast & = \argmin_{\Theta \in \mathcal{R}_\Theta} \sum \limits_{t=0}^{T-1} d_\mathcal{M}\left(x(t+1),\hat{x}(t+1) \right) \\
    & = \argmin_{\Theta \in \mathcal{R}_\Theta} \sum \limits_{t=0}^{T-1} \sum_{i \in \Lambda} \frac{d_{MK}\left((x_i(t+1), \sum \limits_{k=1}^{K} \theta_k \mathcal{N}_i(t) \right)}{2^{\lvert i \rvert}} \nonumber\\
    \text{subject to}  & \quad \sum_{k=1}^K \theta_k = 1, \ 0 \leq \theta_k \leq 1, \forall k \nonumber
\end{align}
\end{small}
where $\hat{x}_i(t+1) = \sum \limits_{k=1}^{K} \theta_k \mathcal{N}_i(t)$; and $\mathcal{R}_\Theta$ gives the parameter search region as follows: $\mathcal{R}_\Theta = \{\theta \in \mathbb{R}^K \mid \sum_{k=1}^K \theta_k = 1, \ 0 \leq \theta_k \leq 1, \forall k\}$. 

\subsection{Differential Evolution based Parameter Estimation Approach}
\label{subsec:PE}

The local rule parameter estimation problem in Eq.~\ref{eq:IP} essentially represents a constrained numerical optimization problem. For such problems, metaheuristic algorithms offer two critical advantages over classical analytical methods: (1) derivative-free optimization eliminates the computational burden of computing gradients, and (2) global search capability through population-based exploration can avoid local optima inherent in multimodal polynomial landscapes. We, therefore, consider an adaptive variant of Differential Evolution (DE), SaDE~\cite{Qin:Huang:2009}. SaDE is a subclass of population-based evolutionary search heuristics, which has been successfully applied to numerical optimization problems across diverse fields. Algorithm~\ref{alg:SaDE_CA_basic} outlines the overall steps involved in the proposed SaDE-based parameter estimation approach. In the following, we briefly discuss the key features of SaDE for the sake of completeness, and we refer to~\cite{Qin:Huang:2009} for a detailed treatment.

SaDE begins with a population $NP$ of candidate vectors, where each candidate represents a possible solution vector, \textit{e.g.}, $\Theta_i = \{ \theta_{i,1}, \theta_{i,2}, \theta_{i,3}, \ldots, \theta_{i,K}\}$. This population is iteratively evolved to identify the \emph{optimal} solution. This is achieved through two evolutionary operations: Mutation (Line~\ref{l:mut1}-\ref{l:mut2}, Algorithm~\ref{alg:SaDE_CA_basic}) and Crossover (Line~\ref{l:xovr1}-\ref{l:xovr2}, Algorithm~\ref{alg:SaDE_CA_basic}). These evolutionary operators generate new trial candidates, denoted by $\tilde{\Theta}$, at each iteration and represent the core of the search operation. A trial candidate can enter the population if it achieves a better \emph{fitness} value (to be discussed later), see Line~\ref{l:J1}-\ref{l:J2}, Algorithm~\ref{alg:SaDE_CA_basic}.

\begin{algorithm}[!t]
\footnotesize
\caption{CA Parameter Identification using SaDE}
\label{alg:SaDE_CA_basic}
\begin{algorithmic}[1]
\STATE Initialize population $\{\Theta_1, \Theta_2, \ldots, \Theta_{NP}\}$
\STATE Initialize adaptive crossover rate $CR$ for each strategy
\STATE \texttt{strategy pool} = \{ \texttt{rand/1}, \texttt{rand-to-best/2}, \texttt{rand/2}, \texttt{current-to-rand/1} \}
\STATE Initialize strategy probabilities:\\
$p_{\texttt{strategy}} = 1/|\texttt{strategy pool}|$ for all $\texttt{strategy} \in \texttt{strategy pool}$
\WHILE{stopping criterion not satisfied}
    \FOR{$i = 1$ to $NP$}
        \STATE Select $\texttt{strategy} \in \texttt{strategy pool}$ using Stochastic Universal Sampling~\cite{Qin:Huang:2009}
        \STATE $k_{rand} = \lfloor \text{rand}[0,1) \times K \rfloor + 1$
        \STATE Random neighbors sampled without replacement: $\{r_1, r_2, r_3, r_4, r_5\} \subset [1,NP] \setminus \{i\}$, 
        \FOR{$k = 1$ to $K$ \nllabel{l:mut1}}
            \IF{\texttt{strategy} = \texttt{rand/1}}
                \STATE $\tilde{\theta}_{i,k} = \theta_{r_1,k} + F(\theta_{r_2,k} - \theta_{r_3,k})$ \nllabel{l:strat1}
            \ELSIF{\texttt{strategy} = \texttt{rand-to-best/2}}
                \STATE $\tilde{\theta}_{i,k} = \theta_{i,k} + F(\theta_{best,k} - \theta_{i,k}) + F(\theta_{r_1,k} - \theta_{r_2,k}) + F(\theta_{r_3,k} - \theta_{r_4,k})$ \nllabel{l:strat2}
            \ELSIF{\texttt{strategy} = \texttt{rand/2}}
                \STATE $\tilde{\theta}_{i,k} = \theta_{r_1,k} + F(\theta_{r_2,k} - \theta_{r_3,k}) + F(\theta_{r_4,k} - \theta_{r_5,k})$ \nllabel{l:strat3} 
            \ELSE 
                \STATE $\tilde{\theta}_{i,k} = \theta_{i,k} + \acute{F}(\theta_{r_1,k} - \theta_{i,k}) + F(\theta_{r_2,k} - \theta_{r_3,k})$ \COMMENT{\texttt{current-to-rand/1}} \nllabel{l:strat4}
            \ENDIF \nllabel{l:mut2}\\
            \IF{$\texttt{strategy} \neq \texttt{current-to-rand/1}$ \nllabel{l:xovr1}}
                \STATE $\tilde{\theta}_{i,k} = \begin{cases}
                    \tilde{\theta}_{i,k}, & \text{if rand}[0,1) < CR \text{ or } k = k_{rand} \\
                    \theta_{i,k}, & \text{otherwise}
                \end{cases}$
            \ENDIF \nllabel{l:xovr2}
        \ENDFOR
        
        \COMMENT{\texttt{Solution Repair}}

        \texttt{bound repair}: $\tilde{\theta}_{i,k}\gets \min\big(\max(\tilde{\theta}_{i,k},0),1 \big)$\\
        
        \texttt{sum-to-one}: $\tilde{\Theta}_i \gets \tilde{\Theta}_i / \sum_{k=1}^{K} \tilde{\theta}_{i,k}$ \nllabel{l:constr2}

        \STATE Evaluate $J(\tilde{\Theta}_i)$ using CA simulation, see Eq.~\ref{eqn:IP_MOP}\nllabel{l:J1}
        \IF{$J(\tilde{\Theta}_i) \leq J(\Theta_i)$}
            \STATE $\Theta_i \gets \tilde{\Theta}_i$, record strategy success
            \IF{$J(\tilde{\Theta}_i) < J(\Theta_{best})$}
                \STATE $\Theta_{best} \gets \tilde{\Theta}_i$
            \ENDIF
        \ELSE
            \STATE Record strategy failure
        \ENDIF \nllabel{l:J2}
    \ENDFOR
    \STATE Update strategy probabilities and $CR$ based on success/failure rates, see~\cite{Qin:Huang:2009}
\ENDWHILE
\RETURN $\Theta_{best}$
\end{algorithmic}
\end{algorithm}

It is worth highlighting that the \emph{mutation} operation explores new trial candidates by sampling the search region $\mathcal{R}_\Theta$, and, therefore, is crucial to successful optimization. For instance, consider a well-known \texttt{rand/1} mutation strategy, which combines three randomly selected candidates from the population to generate a new trial candidate, as follows:
\begin{equation}
    \label{eq:rand/1}
    \tilde{\theta}_{i,k} = \theta_{r_1,k} + F \cdot (\theta_{r_2,k} - \theta_{r_3,k})
\end{equation}
where, $r_1,r_2,r_3 \in \llbracket 1,NP \rrbracket$ denote randomly selected candidates from the population; $F$ is the scaling factor. It is known that \texttt{rand/1} strategy demonstrates excellent \emph{exploration} capability albeit at the expense of slower convergence; such qualities are desirable in multi-modal search regions. Significant efforts have been dedicated to developing different mutation strategies to better adapt to different features of the search region (\textit{e.g.}, uni-modal or multi-modal). However, the nature of the problem landscape is often not known \emph{a priori}, so the selection of the appropriate mutation strategy can be challenging. SaDE~\cite{Qin:Huang:2009} was developed to address such issues by maintaining a pool of four strategies, which are probabilistically selected based on their performance over a fixed number of past iterations. Algorithm~\ref{alg:SaDE_CA_basic} shows these mutation strategies:
\texttt{rand/1} (Line~\ref{l:strat1}), \texttt{rand-to-best/2} (Line~\ref{l:strat2}), \texttt{rand/2} (Line~\ref{l:strat3}) and \texttt{current-to-rand/1} (Line~\ref{l:strat4}). We refer to~\cite{Qin:Huang:2009} for a detailed discussion on the capabilities of each mutation strategy, the probabilistic update mechanisms of mutation strategies, and the Crossover Rate ($CR$).

We consider a \emph{solution-repair} approach (see Line~\ref{l:constr2}, Algorithm~\ref{alg:SaDE_CA_basic}) to ensure that all trial candidates are \emph{feasible} and satisfy the equality constraint: $\sum_k \theta_k = 1$, see Eq.~\ref{eq:IP}. The \emph{fitness} of each trial candidate is calculated next using the distance $d_{MK}$, as follows:
\begin{small}    
\begin{align}
\label{eqn:IP_MOP}
J(\tilde{\Theta}_i) & =\sum \limits_{t=0}^{T-1} \sum_{i \in \Lambda} \frac{d_{MK}\left((x_i(t+1), \sum \limits_{k=1}^{K} \tilde{\theta}_{i,k} \mathcal{N}_i(t) \right)}{2^{\lvert i \rvert}} 
\end{align} %
\end{small}
Further, drawing on the recommendations in~\cite{Qin:Huang:2009}, the search parameters of SaDE are set as follows: population size, $NP\gets100$; learning period, $LP\gets50$; scaling factors 
$F$ which are sampled from a normal distribution with $\text{mean}=0.5$ and $\text{variance}=0.3$, adaptive crossover rate $CR$ (initial value $0.5$), and four strategies with equal initial probabilities $p_{\texttt{strategy}} = 0.25$. A total of 20 independent runs of SaDE are carried out, where each run is terminated upon reaching the maximum number of iterations ($G=500$) to account for the stochastic nature of the algorithm.

\section{Results}
\label{sec:Res}
\subsection{Test Scenarios}
\label{subsec:testscene}

\begin{figure}[!b]
    \centering
    \begin{tabular}{cccc}
        \includegraphics[width=0.25\textwidth]{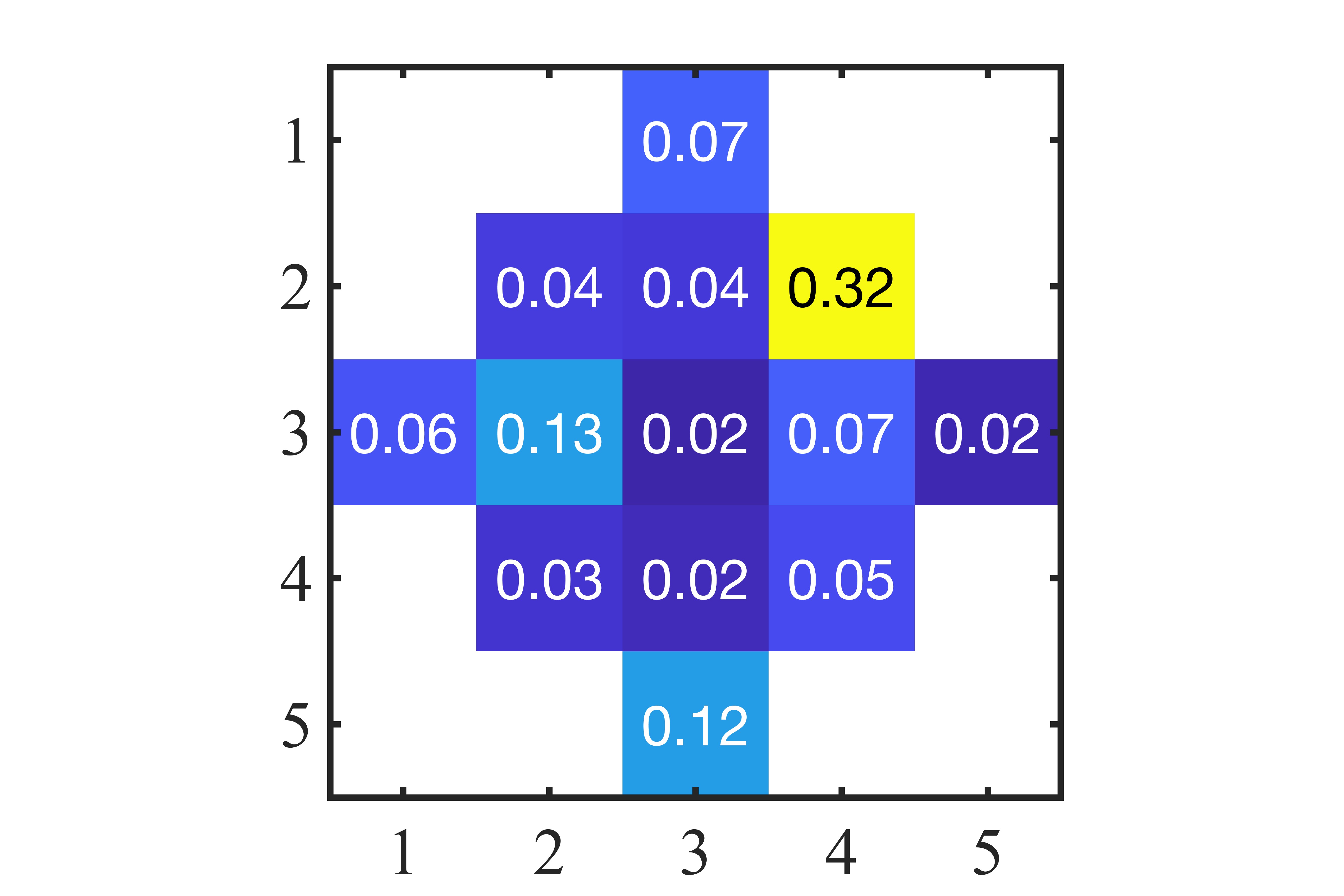} & 
         \includegraphics[width=0.25\textwidth]{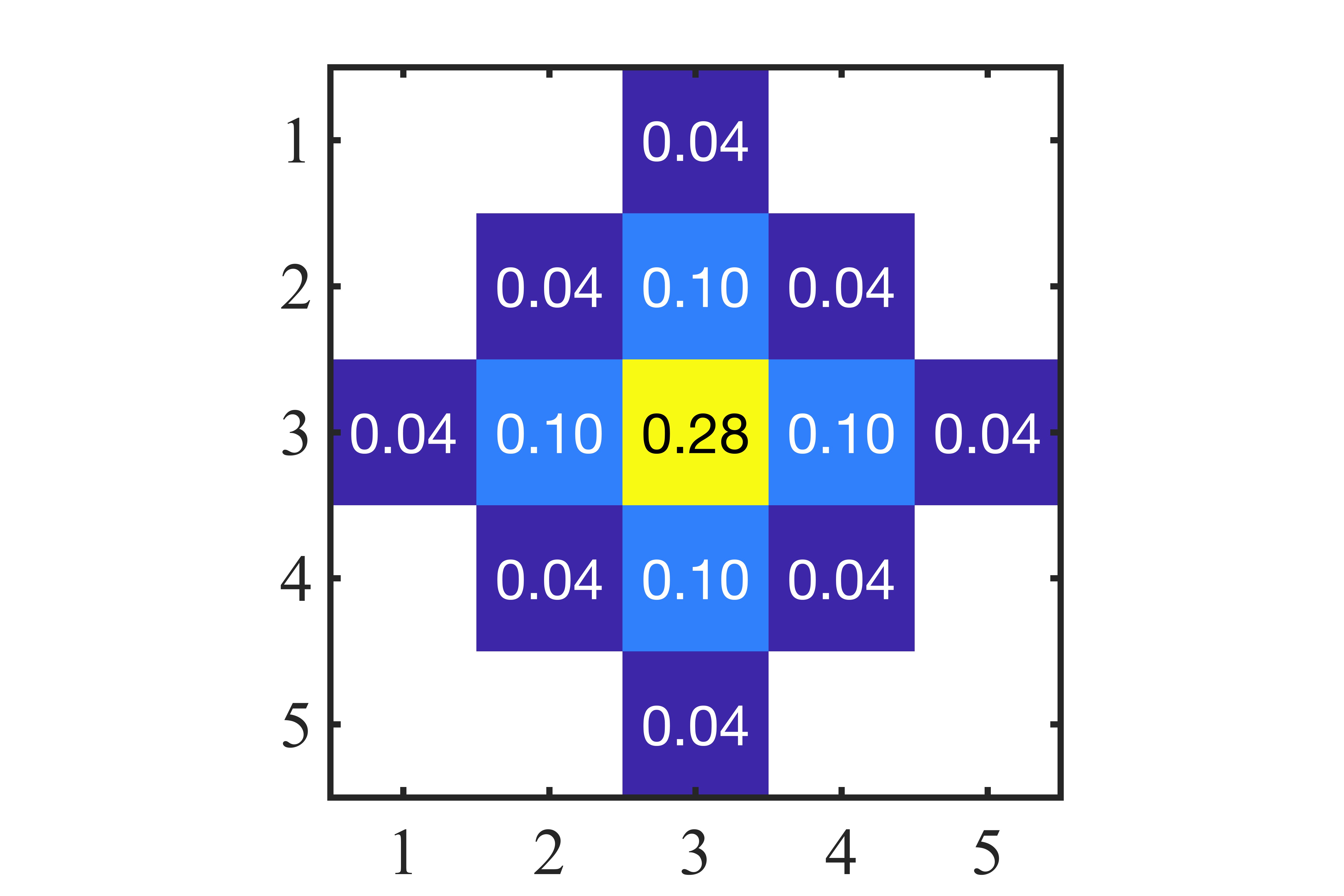} &
        \includegraphics[width=0.25\textwidth]{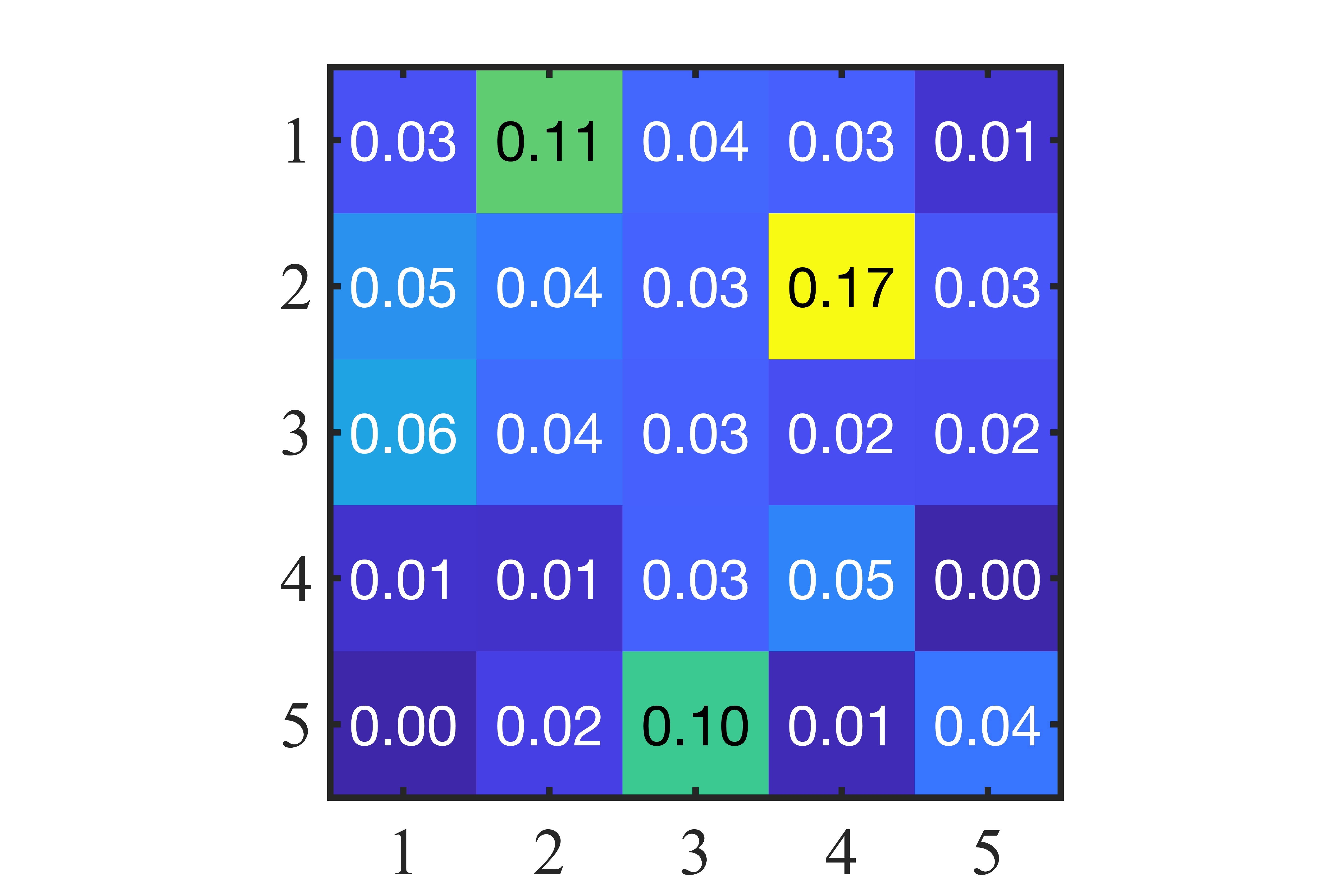} &
       \includegraphics[width=0.25\textwidth]{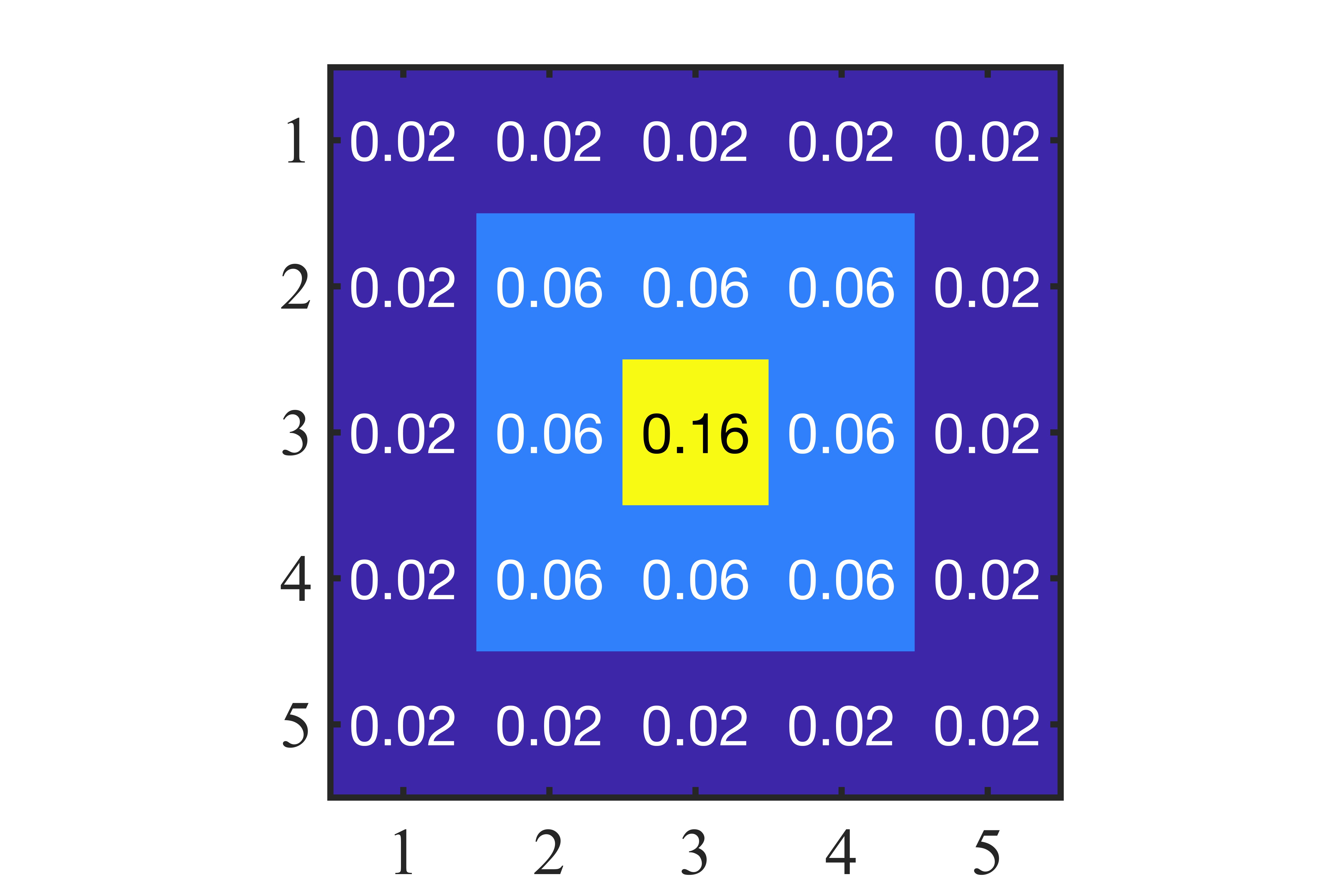}\\
        (a)\hspace{1mm} \makecell{Manhattan\\ + \emph{random}} &   
        (b)\hspace{1mm} \makecell{Manhattan\\ + \emph{distance-based}} &
        \hspace{0.2mm} (c) \makecell{Moore\\ + \emph{random}} &   
        (d)\hspace{1mm} \makecell{Moore\\ + \emph{distance-based}}\\
    \end{tabular}
    \caption{Parameter assignment for $r=2$.}
    \label{fig:Casetheta}
\end{figure}

We consider several identification scenarios to emulate distinct local behaviors. This is achieved by variations in the local neighborhood of a two-dimensional CAM: \textit{two neighborhood types} (Manhattan/von Neumann and Moore) and \textit{three neighborhood radii} ($r=1,2,$ and $3$). Moreover, two sets of parameters are generated for each neighborhood: the first set is generated \emph{randomly}, \textit{i.e.}, each neighbor is assigned a random parameter $\theta_k \in [0,1]$, see Fig.~\ref{fig:Casetheta}a~and~~\ref{fig:Casetheta}c. In contrast, the other set (denoted by \emph{distance-based}) assigns parameter values that are inversely proportional to distance, $\theta_k$, see Fig.~\ref{fig:Casetheta}b~and~~\ref{fig:Casetheta}d. This leads to a total of 12 different identification scenarios (2 neighborhood types $\times$ 3 radii $\times$ 2 parameter sets). For convenience, we split these scenarios into two cases: \emph{Case-I} includes all the scenarios where $\theta_k$ values are randomly assigned, whereas \emph{Case-II} encompasses the remaining scenarios. Note that the parameter assignments under both cases ensure the sum of parameters $\sum_k \theta_k = 1$.
\begin{figure}[!t]
    \centering
    \begin{tabular}{c}
        \includegraphics[width=\textwidth]{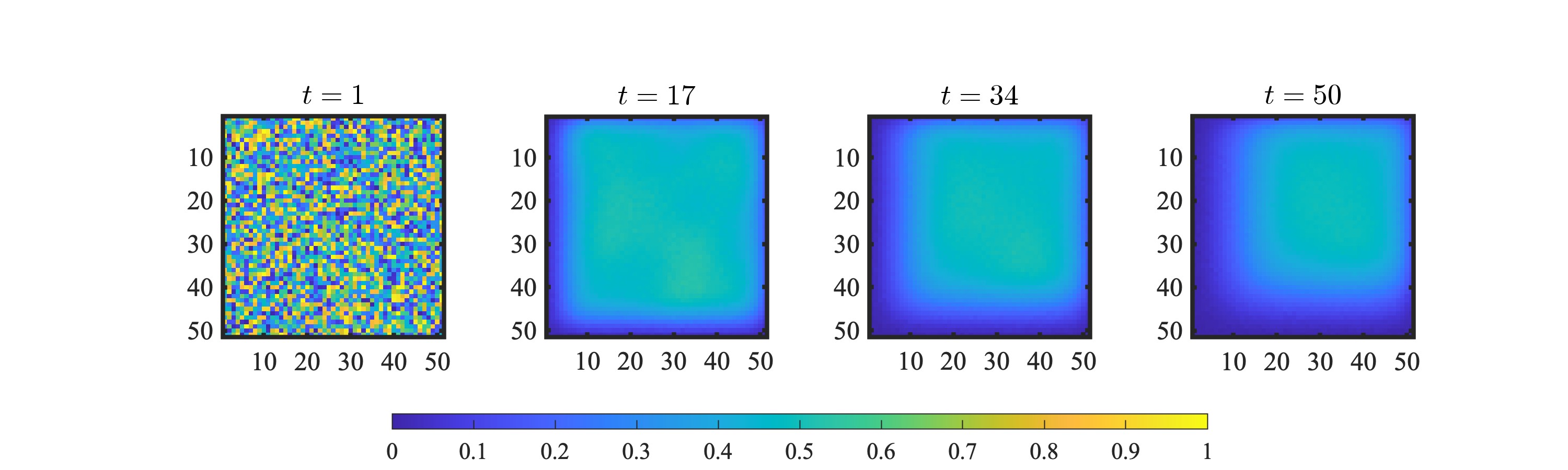}\\
        (a)\hspace{1mm} Random $\theta$ assignment\\   
        \includegraphics[width=\textwidth]{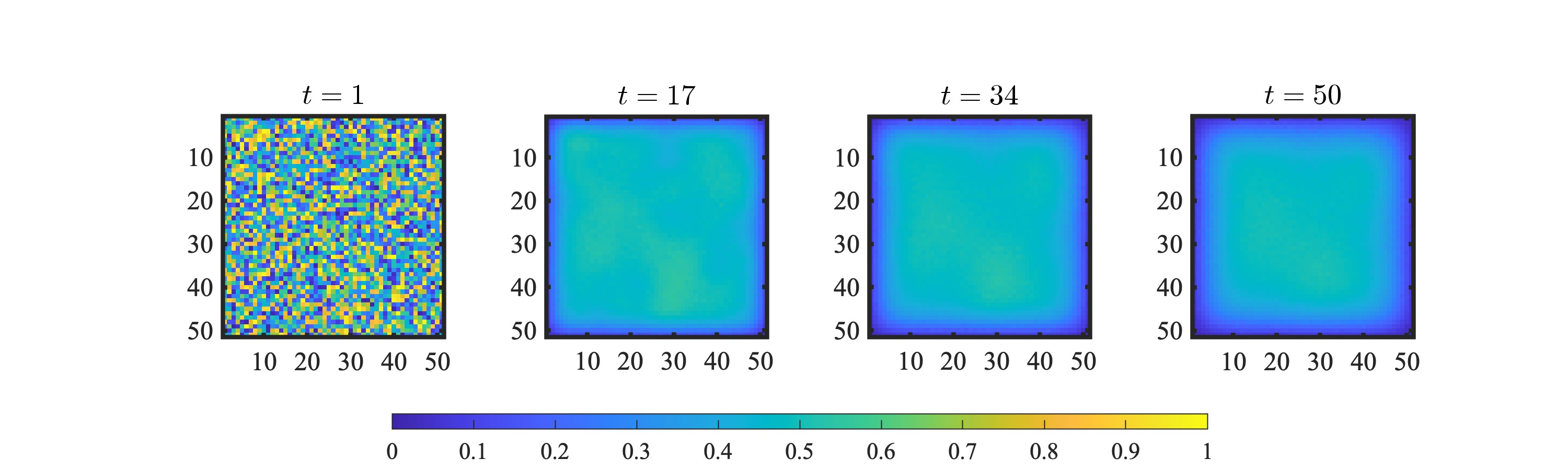}\\
        (b)\hspace{1mm} Distance-based $\theta$ assignment\\   
    \end{tabular}
    \caption{Evolution of CAMs with Manhattan neighborhood and with $r = 2$}
    \label{fig:CAMEvol}
\end{figure}

In each scenario, the identification data $\mathcal{O}$ is collected on a two-dimensional rectangular lattice ($\Lambda = 51 \times 51$ cells) over $T=10$ time steps. Note that this temporal horizon is selected through trial-and-error to balance computational complexity with parameter identifiability. A Gaussian white noise ($SNR = 40 \ dB$) is added to the observed states to simulate the real-world measurement noise. Each scenario begins with the same randomly assigned $\in[0,1]$ initial states at $T=0$, which ensures that any change in dynamic behavior across multiple scenarios can be ascribed to the local rule. This is further illustrated in Fig.~\ref{fig:CAMEvol}, which demonstrates CAM evolution with the Manhattan topology ($r=2$) over four different time steps under two scenarios: Fig.~\ref{fig:CAMEvol}a with \emph{randomly} assigned parameters shown in Fig.~\ref{fig:Casetheta}a, and Fig.~\ref{fig:CAMEvol}b with \emph{distance-based} parameters shown in Fig.~\ref{fig:Casetheta}b.

\subsection{Parameter Estimation Results}
\label{subsec:resParaEst}

Given that the identification data is collected using the known local rule (and thus the parameter $\theta_k$), we use the Normalized Root Mean Square Error (NRMSE) metric to gauge the accuracy of the estimates $\hat{\theta}_k$, as follows:
\begin{small}
\begin{equation}
\label{eq:TRE}
     NRMSE = \frac{\sqrt{\frac{1}{K}\sum_{k=1}^{K}(\hat{\theta}_k - \theta_k)^2}}{\max(\Theta) - \min(\Theta)} \times 100
\end{equation}
\end{small}

It is worth noting that the equality constraint, $\sum_k \theta_k = 1$, dictates that overall parameter values decrease with the increase in the neighborhood radius,  which can skew error metrics. NRMSE avoids numerical instability with near-zero parameters in such scenarios and provides consistent comparisons across different neighborhood sizes and types. 

\begin{figure}[!t]
    \centering
    \begin{tabular}{cc}
        \includegraphics[width=0.48\textwidth]{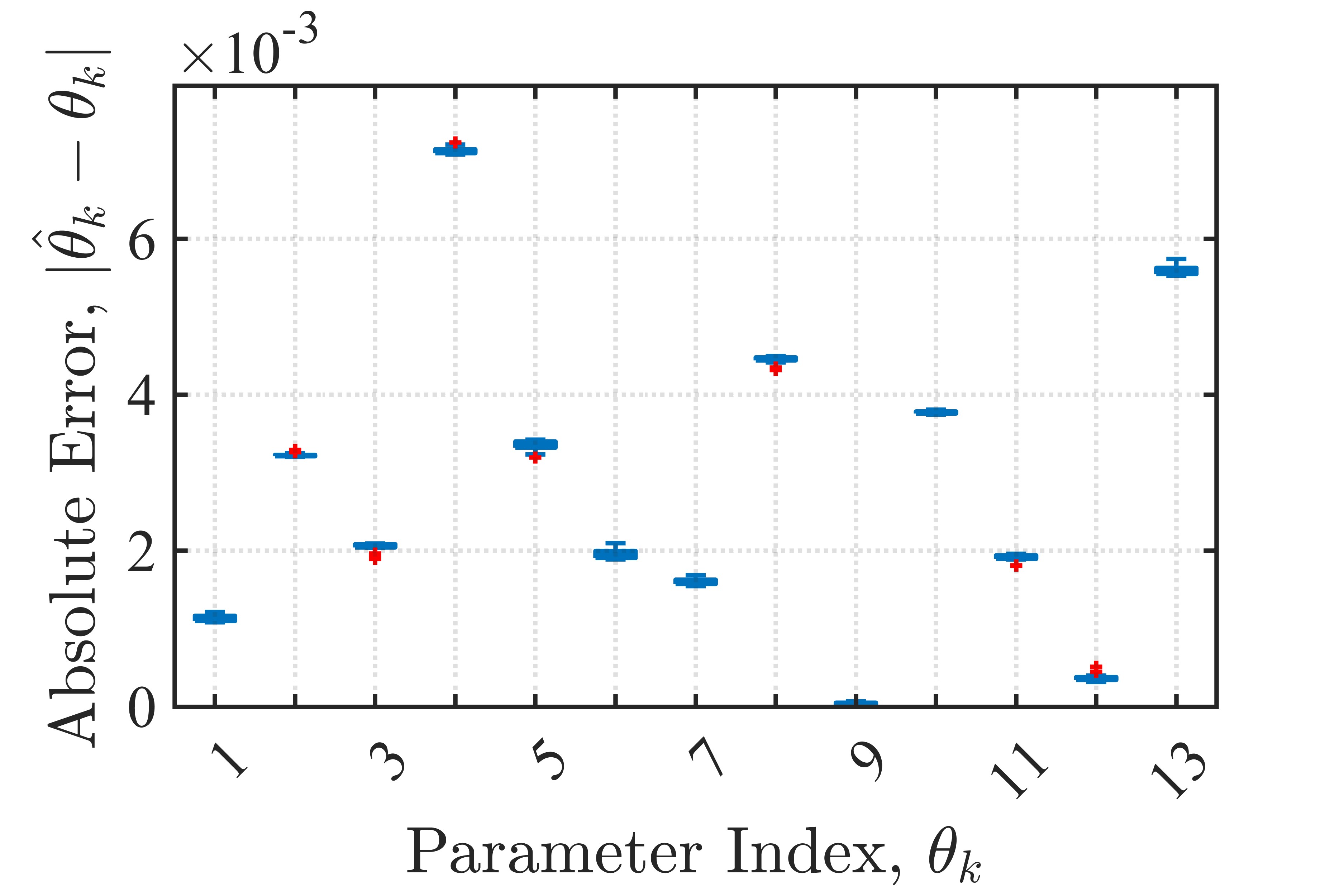}
        &
        \includegraphics[width=0.48\textwidth]{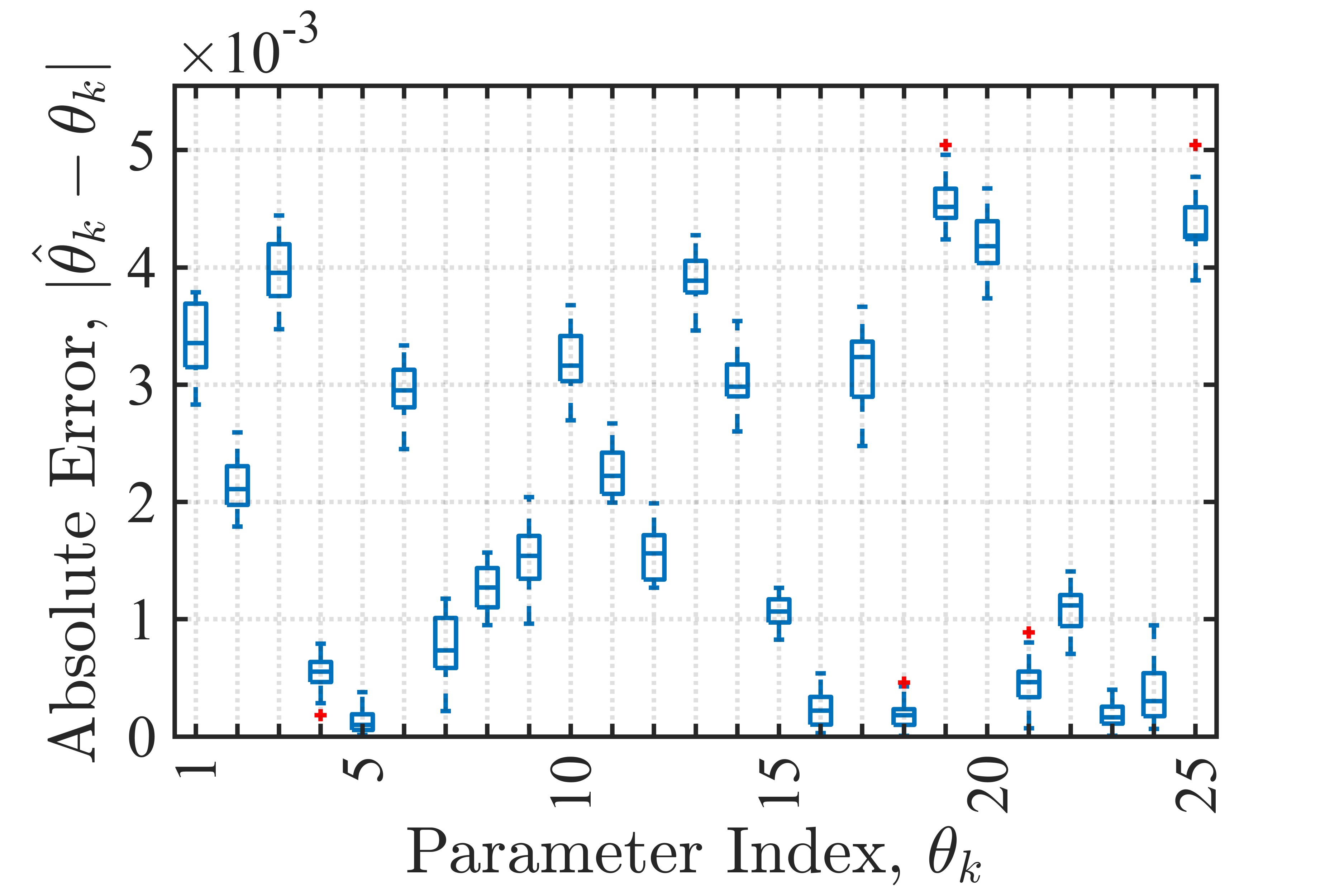}\\
        (a)\hspace{1mm} Manhattan ($r=2, K=13$) & (b)\hspace{1mm} Moore ($r=2, K=25$)
    \end{tabular}
    \caption{Parameter Estimation Error with random $\theta$ assignments}
    \label{fig:barplotrandom}
\end{figure}

To evaluate the efficacy of the proposed approach, the parameter estimation is carried out on all identification scenarios (Section~\ref{subsec:testscene}) following the search setup described in Section~\ref{subsec:PE}. A total of 20 independent runs of SaDE are carried out for each identification scenario due to its stochastic nature. At the end of each run, the estimated parameters, $\hat{\Theta} = \{ \hat{\theta}_1, \hat{\theta}_2, \ldots, \hat{\theta}_K \}$, are recorded for further analysis. Fig.\ref{fig:barplotrandom} depicts such variations in the absolute parameter estimation error, $|\hat{\theta}_k - \theta_k|$, over 20 independent runs of SaDE for the identification scenarios with $r=2$ and random parameter assignments. 

For the sake of brevity, we summarize the parameter estimation results of all identification scenarios over 20 independent runs of SaDE in terms of NRMSE in Table~\ref{t:ParResCombined}. The corresponding variations in the fitness function, $J(\cdot)$, are also shown in Table~\ref{t:JResCombined}. The results indicate an excellent parameter recovery with $\text{NRMSE} \leq 2\%$ across all identification scenarios. As expected, the neighborhood size impacts the estimation performance; the quadratic increase in parameter count ($K$) with increasing radius ($r$) slightly affects parameter recovery. For example, the mean NRMSE values of Manhattan neighborhoods rise from~0.8-1.3\% (with $r=1$) to~1.7-2.2\% (with $r=3$). 

Moreover, the fitness function results in Table~\ref{t:JResCombined} further demonstrate that the proposed approach can accurately capture spatio-temporal dynamics. For all identification scenarios, the average value of the fitness function $J(\cdot) \equiv d_\mathcal{M} \leq 0.22$, which indicates the predicted states $\hat{x}$ closely simulate the observed states $x$ (see Eq.~\ref{eq:IP}~and~\ref{eqn:IP_MOP}), even in the presence of measurement noise. Further, the results show consistently better performance for the Manhattan neighborhoods irrespective of the neighborhood radius, $r$. This, in part, can be explained by its relatively compact structure, which leads to fewer parameters ($K$) for a particular radius compared to the Moore neighborhood. Additionally, distance-based parameter assignments yield marginally better fitness values than random assignments, particularly for Manhattan neighborhood, indicating that parameter distribution patterns influence identification performance.

\begin{table*}[!t]
  \centering
  \caption{Parameter estimation errors over 20 independent runs of SaDE}
  \label{t:ParResCombined}%
  \begin{adjustbox}{width=0.75\textwidth}
  \begin{threeparttable}
    \begin{tabular}{c|cc|cc|cc}
    \toprule
    \multirow{2}[4]{*}{\textbf{NRMSE}} & \multicolumn{2}{c|}{\boldmath $r=1$} & \multicolumn{2}{c|}{\boldmath $r=2$} & \multicolumn{2}{c}{\boldmath $r=3$} \\
    \cmidrule{2-7}          & \makecell{\textbf{Manhattan,}\\ \boldmath $K=5$} & \makecell{\textbf{Moore,}\\ \boldmath $K=9$} & \makecell{\textbf{Manhattan,}\\ \boldmath $K=13$} & \makecell{\textbf{Moore,}\\ \boldmath $K=25$} & \makecell{\textbf{Manhattan,}\\ \boldmath $K=25$} & \makecell{\textbf{Moore,}\\ \boldmath $K=49$} \\
    \midrule
    \multicolumn{7}{c}{\textit{Case-I: Random $\theta$ Assignment}} \\
    \midrule

    \textbf{Best} & 0.784 & 0.497 & 1.123 & 1.451 & 1.676 & 1.722 \\
    \textbf{Mean} & 0.784 & 0.497 & 1.128 & 1.548 & 1.736 & 2.042 \\
    \textbf{SD} & 5.52E-11 & 5.70E-06 & 3.05E-03 & 4.75E-02 & 3.92E-02 & 2.01E-01 \\
    \textbf{Max} & 0.784 & 0.497 & 1.133 & 1.631 & 1.830 & 2.396 \\

    \midrule
    \multicolumn{7}{c}{\textit{Case-II: Distance-based $\theta$ Assignment}} \\
    \midrule

    \textbf{Best} & 1.314 & 1.301 & 1.603 & 1.774 & 1.947 & 1.774 \\
    \textbf{Mean} & 1.314 & 1.301 & 1.605 & 1.861 & 2.185 & 1.972 \\
    \textbf{SD} & 6.85E-12 & 5.25E-06 & 2.29E-03 & 5.02E-02 & 1.22E-01 & 9.03E-02 \\
    \textbf{Max} & 1.314 & 1.301 & 1.612 & 1.948 & 2.362 & 2.100 \\

    \bottomrule
    \end{tabular}%
  \end{threeparttable}
 \end{adjustbox}
\end{table*}
\begin{table*}[!t]
  \centering
  \caption{Variations in the fitness function $J(\cdot)$ over 20 independent runs of SaDE}
  \label{t:JResCombined}%
  \begin{adjustbox}{width=0.75\textwidth}
  \begin{threeparttable}
    \begin{tabular}{c|cc|cc|cc}
    \toprule
    \multirow{2}[4]{*}{\boldmath$J(\Theta_{best})$} & \multicolumn{2}{c|}{\boldmath $r=1$} & \multicolumn{2}{c|}{\boldmath $r=2$} & \multicolumn{2}{c}{\boldmath $r=3$} \\
    \cmidrule{2-7}          & \makecell{\textbf{Manhattan,}\\ \boldmath $K=5$} & \makecell{\textbf{Moore,}\\ \boldmath $K=9$} & \makecell{\textbf{Manhattan,}\\ \boldmath $K=13$} & \makecell{\textbf{Moore,}\\ \boldmath $K=25$} & \makecell{\textbf{Manhattan,}\\ \boldmath $K=25$} & \makecell{\textbf{Moore,}\\ \boldmath $K=49$} \\
    \midrule
    \multicolumn{7}{c}{\textit{Case-I: Random $\theta$ Assignment}} \\
    \midrule

    \textbf{Best} & 0.16  & 0.27  & 0.12  & 0.23  & 0.14  & 0.21 \\
    \textbf{Mean} & 0.16  & 0.27  & 0.12  & 0.23  & 0.14  & 0.21 \\
    \textbf{SD} & 1.13E-15 & 7.08E-09 & 1.31E-06 & 2.01E-05 & 7.17E-05 & 1.00E-03 \\
    \textbf{Max} & 0.16  & 0.27  & 0.12  & 0.23  & 0.14  & 0.22 \\
   
    \midrule
    \multicolumn{7}{c}{\textit{Case-II: Distance-based $\theta$ Assignment}} \\
    \midrule

    \textbf{Best} & 0.14  & 0.26  & 0.12  & 0.24  & 0.12  & 0.22 \\
    \textbf{Mean} & 0.14  & 0.26  & 0.12  & 0.24  & 0.12  & 0.22 \\
    \textbf{SD} & 8.52E-17 & 3.56E-09 & 1.45E-06 & 4.41E-05 & 1.73E-04 & 4.75E-04 \\
    \textbf{Max} & 0.14  & 0.26  & 0.12  & 0.24  & 0.12  & 0.22 \\
      
    \bottomrule
    \end{tabular}%
  \end{threeparttable}
 \end{adjustbox}
\end{table*}
\subsection{SaDE: Convergence and Runtime Analysis}
\label{subsec:resRuntime}

\begin{figure}[!t]
    \centering
    \begin{tabular}{cc}
      \includegraphics[width=0.48\textwidth]{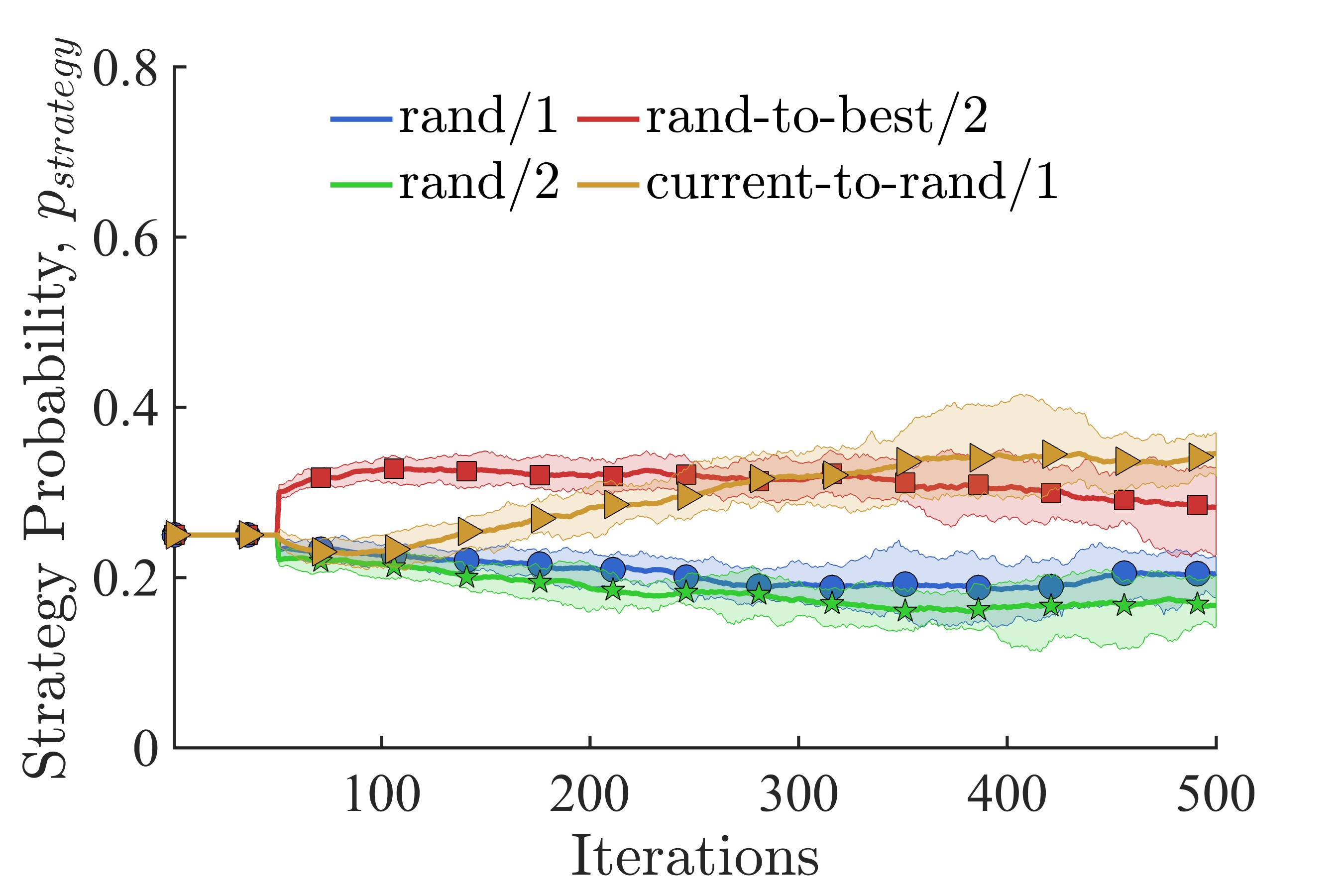} & \includegraphics[width=0.48\textwidth]{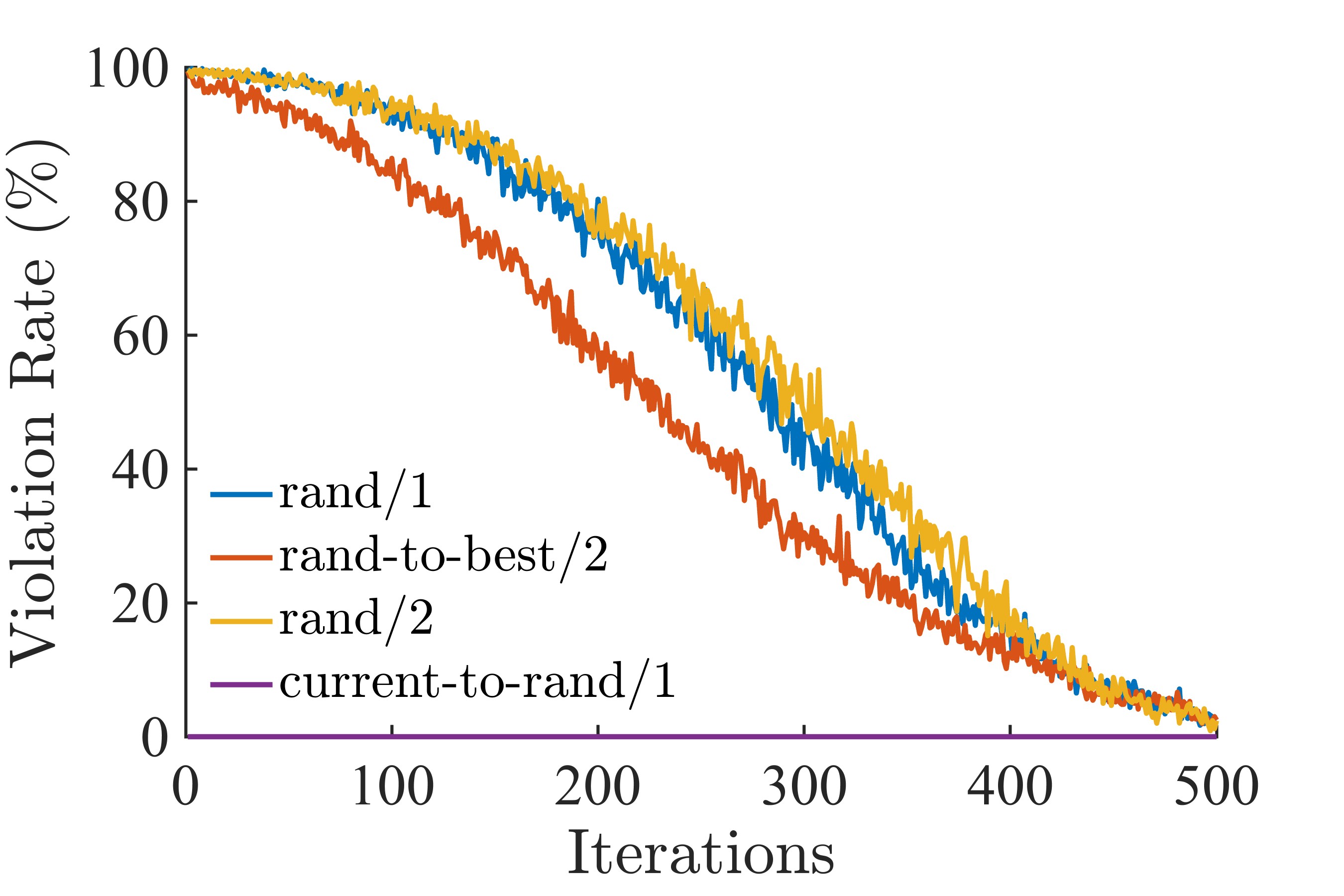} 
     \\[2mm]
    (a)\hspace{1mm} Iterative Strategy Probability & \hspace{0.5mm} (b)\hspace{1mm} Iterative Infeasible Solution Rate
    \end{tabular}
    \caption{Comparison of SaDE mutation strategies over 20 independent runs on Manhattan topology with $r=3$ and \emph{random} $\theta$ assignment.}
    \label{f:runtimeanalysis}
\end{figure}

Finally, we focus on the mutation strategies of SaDE and their impacts on the proposed parameter estimation problem. To this end, we observed the iterative variations in the selection probabilities, $p_{\texttt{strategy}}$, of all strategies being considered: $\{\texttt{rand/1}, \texttt{rand-to-best/2}, \texttt{rand/2}, \texttt{current-to-rand/1}\}$, see Algorithm~\ref{alg:SaDE_CA_basic}. Fig.~\ref{f:runtimeanalysis}a depicts these results for the Manhattan topology with $r=3$ and \emph{random} $\theta$ assignments. The results clearly demonstrate that two strategies dominate throughout the search process: \texttt{rand-to-best/2} and \texttt{current-to-rand/1}. Similar patterns are also observed in the remaining identification scenarios (not shown here for brevity). This may be ascribed to a key challenge of the proposed parameter estimation problem:  the equality constraint, $\sum_k \theta_k =1$. The evolutionary operators (such as mutation and crossover) associated with a particular strategy are likely to generate \emph{infeasible} solutions that violate this constraint.

It is worth emphasizing that both \emph{dominant} strategies use the current parameter vector $\theta_{i,k}$ as their \emph{mutation} basis to generate the trial candidate, see Line~\ref{l:strat2} (\texttt{rand-to-best/2}) and Line~\ref{l:strat4} (\texttt{current-to-rand/1}), Algorithm~\ref{alg:SaDE_CA_basic}. This key distinction is likely to reduce the disruption arising from the binomial crossover operation (see Line~\ref{l:xovr1}-\ref{l:xovr2}, Algorithm~\ref{alg:SaDE_CA_basic}), and thereby conserve parameter relationships and limit constraint violations. Additionally, \texttt{current-to-rand/1} replaces binomial crossover with an arithmetic recombination, possibly leading to reductions in infeasible solutions. To test this hypothesis, in each iteration, we calculated the ratio between infeasible and total solutions (denoted as the Constraint Violation Rate) generated by a particular strategy, before the solution repair is applied (see Line~\ref{l:constr2}, Algorithm~\ref{alg:SaDE_CA_basic}). Fig.~\ref{f:runtimeanalysis}b depicts the results of this analysis, which confirm the hypothesis: \texttt{current-to-rand/1} maintains near-zero violations throughout, while \texttt{rand-to-best/2} reduces violations faster than the remaining strategies.

\begin{figure}[!t]
    \centering
    \begin{tabular}{cc}
    \includegraphics[width=0.48\textwidth]{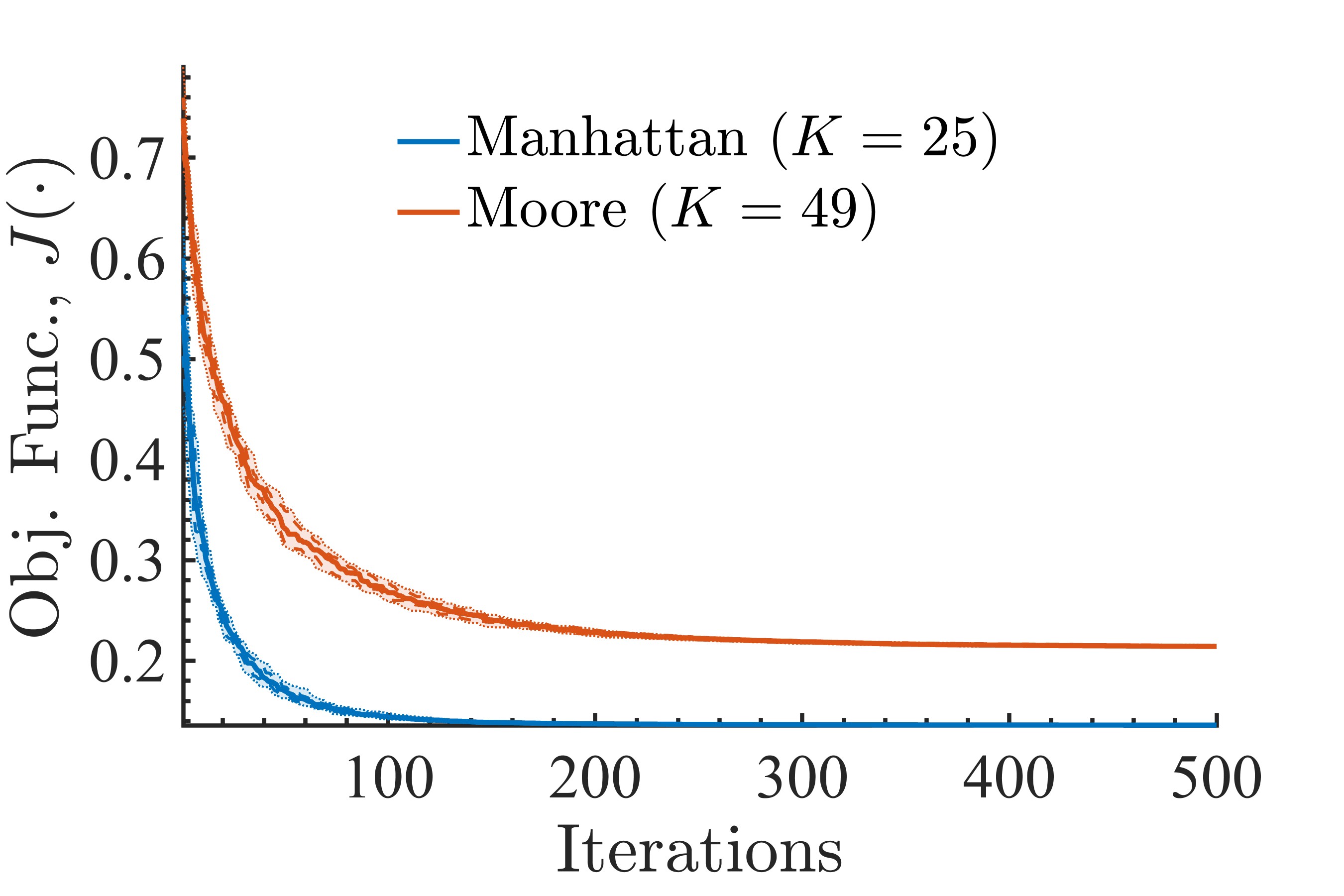} & \includegraphics[width=0.48\textwidth]{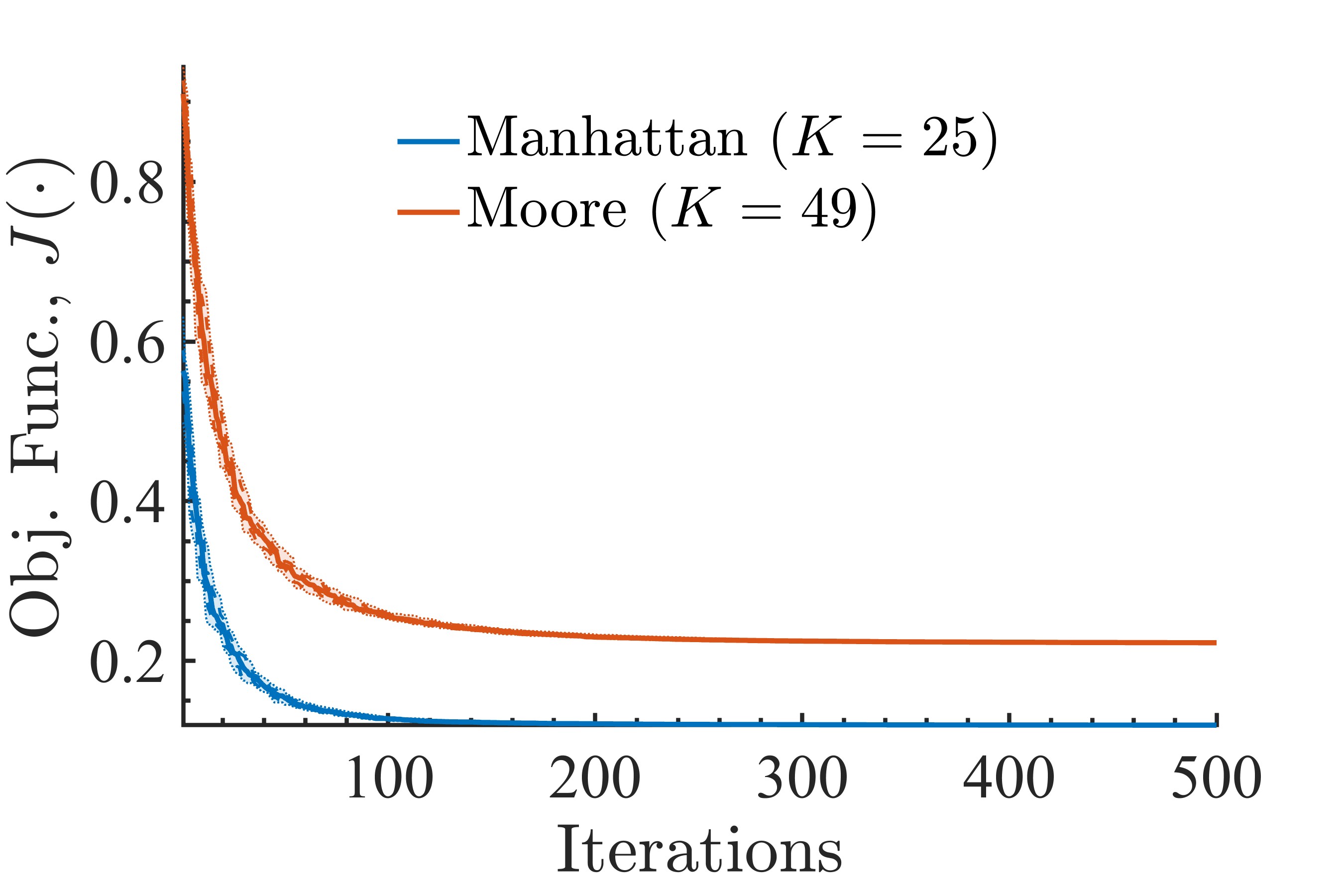} 
     \\[2mm]
    (a)\hspace{1mm} random assignment & (b)\hspace{1mm} distance-based assignment
    \end{tabular}
    \caption{Convergence behavior of SaDE over 20 independent runs for different topologies with $r=3$. Note that y-axis is zoomed and begin from $J(\cdot) \approx 0.14$ to appreciate the difference in $J(\cdot)$ obtained on Manhattan and Moore topologies.}
    \label{f:convDist}
\end{figure}

\section{Conclusions}
\label{sec:conclusions}

A new approach to identifying local rules from the observed data of Cellular Automata on spaces probability measures (CAMs) was proposed. To this end, a special class of CAMs was considered, where cell states are represented by Bernoulli probability distributions and local rules act on such probabilistic configurations. We leveraged the fact that the local rules can be formulated as a convex combination of neighborhood states in this framework. Building upon this notion, it was demonstrated that the identification can be approached as a constrained optimization task involving the estimation of local rule parameters, assuming that the neighborhood size and type are \emph{a priori} determined. Next, a meta-heuristic identification approach was proposed that can estimate the local rule parameters directly from the observed CAM states using Self-adaptive Differential Evolution (SaDE). The performance of the proposed approach was evaluated by considering 12 distinct identification scenarios that involved varied combinations of neighborhood types, neighborhood radii and parameter assignments. The results demonstrate that the known local rule parameters of the test scenarios could be accurately estimated even in the presence of measurement noise. These results are encouraging and demonstrate the potential of the proposed approach to accurately identify local rules governing the observed spatio-temporal data of real-world systems. 

Finally, while this study assumes the \emph{a priori} knowledge of neighborhood size and type, this issue remains one of the fundamental identification challenges, and it will be the topic of our further investigation. Furthermore, the runtime analysis of SaDE underlines the equality constraint, $\sum_k \theta_k =1$, as the key optimization challenge. In this context, some of the existing mutation and crossover strategies were found to be disruptive and led to infeasible solutions. A further investigation into the design or selection of new evolutionary operators to address such issues seems promising.

%
 \bibliographystyle{splncs04}
 \bibliography{article}

\end{document}